\documentclass[%
 aip,
jcp,
 amsmath,amssymb,
preprint,
]{revtex4-2}

\usepackage{graphicx}
\usepackage{dcolumn}
\usepackage{bm}

\usepackage[utf8]{inputenc}
\usepackage[T1]{fontenc}
\usepackage{mathptmx}
\usepackage{etoolbox}

\begin{document}


\title{Kramers rate for Brownian particles in excitable media with deformable double-well substrates}
 
\author{Alain M. Dikand\'e}
\email{dikande.alain@ubuea.cm}

\affiliation{Laboratory of Research on Advanced Materials and Nonlinear Sciences (LaRAMaNS),
Department of Physics, Faculty of Science, University of Buea P.O. Box 63 Buea, Cameroon}%
             
\begin{abstract}
 We address the Kramers escape problem for Brownian particles in bistable substrates with  deformable double-well shapes. The shape deformability is considered of three distinct forms: in one, the positions of the two degenerate minima can be shifted continuously without affecting the barrier height. In the second the minima positions are kept fix while the barrier height is continuously shifted. In the third the minima positions and barrier height can be tuned simultaneously by changing a single real parameter that we refer to as deformability parameter. We obtain the analytical expression of the Kramers escape rate for the three different double-well models, and identify a critical value of the deformability parameter above which non-Gaussian corrections become relevant. Remarkable enough in the latter region, statistical mechanics predicts a first-order transition in quantum tunneling using the functional action associated with the family of deformable double-well potentials.
\end{abstract}  

\keywords{Bistable systems, deformable double-well potential, Kramers escape rate, non-Gaussian corrections, parabolic cylinder function.}

\maketitle
\section{Introduction}
Brownian motors are active small-size particles and microorganisms such as enzymes, bacteria, fungi, protozoa, algae and biochemical agents living in noisy excitable media where the background substrate, created by interactions with their environments, exhibits a well-defined spatial symmetry. Although the majority of models proposed to investigate the dynamics of Brownian motors deals with excitable media with periodic substrates (i.e. the sine-Gordon or double sine-Gordon substrates)  \cite{hang1,hang2,hang3,hang4,hang5,hang6,hang7,dsg1,dsg2,dsg3}, it is well known that most phenomena occuring in biophysical systems resulting from the spatiotemporal organizations of microorganisms involve two-state processes \cite{x1,x2}. This observation is all the more relevant as it attests to the fundamental role bistability plays in the spatial and temporal evolutions of states of biophysical systems, due to a double-well symmetric substrate caused by the interaction of microorganisms with their environments. \par Brownian motors are generally considered to be biochemical machines whose role is to regulate the execution of vital biophysical functions, moreover they interfer in several contexts of two-state processes occuring in biophysics and biochemistry. For instance Brownian motors are known to control the bistable responses of Escherichia coli lac operon intake \cite{bi1,bi2,bi3,bi33,bi34}, they serve as self-regulator of synthetic biochemical activities in biological systems \cite{bi4}, they regulate intracellular transport and signaling in cells and tissues \cite{bi5}. A Brownian motor living in protein microtubules and called Kinesin \cite{bi6,bi7}, has been shown to regulate intracellular energy transfers while standing for the triggerer of conformational transitions in macrotubular proteins \cite{bi6,bi8}. Brownian motors regulate catalytic rates in enzymes, molecular proteins \cite{bi8a} and so on. Stress that in biophysical systems in general and DNA in particular, conformational changes take place the same way structural transitions occur in solids and some molecular crystals. In these systems the change in cell conformations involves two metastable states corresponding to the unstable and stable conformations, in the Ginzburg-Landau picture of structural transitions the two conformations are isoenergetic and are separated by a finite energy barrier  \cite{a1,a2,gl1,gl2,gl3,gl4,gl5,gl6}.  \par
The dynamics of Brownian particles in excitable media with double-well potentials has been investigated in the general context of physical systems with substrates characterized by a bistable symmetry \cite{cof1,cof2,cof3}. The associated problem of metastable-state decays has also attracted much attention, given its importance in thermally activated kinetic processes. Pioneer in the analysis of this problem Kramers was originally interested in the velocity of chemical reactions \cite{1,7}. Indeed considering a particle in an arbitrary potential with eventually one or several wells separated by barriers, Kramers obtained that the escape probability of the particle in the arbitrary potential interacting with a heat bath of equilibrium fluctuations, exhibits an exponential signature typical of Arrhenius' activation except a prefactor that depends on wether the interaction of the particle to the substrate potential in the presence of the heat bath is moderate-to-strong, or weak. Kramers' rate theory has been extensively discussed in connection with biophysical processses \cite{99}, chemical reactions \cite{12,13} and randomly distributed oscillators in statistical physics \cite{7,5,6,8,98}. \par In most of the aforementioned studies of escape-rate problem in bistable systems, the model considered rests on the assumption that the Brownian particle moves in the field of force created by the Ginzburg-Landau free energy quite familiar in the literature of structural transitions of second order. This free energy can also be represented in a canonical form known as $\phi^4$ potential \cite{18a,18b,18c,18d,18e} which writes:
\begin{equation}
V(x)=\frac{a_0}{8}(x^2-1)^2. \label{eq1}
\end{equation}
Equation (\ref{eq1}) features a double-well potential with two degenerate minima $x=\pm 1$, and a maximum $x=0$ where a barrier of height $V(0)=a_0/8$ is erected ($a_0$ is a real and positive constant). While the existence of two minima and a finite potential barrier agrees with observations on real bistable systems, the fixed minima positions of the $\phi^4$ potential is a weakness since in real contexts and particularly in biophysical systems the double-well shape of the substrate can be affected by several physical factors. Isotopic and solvent effects, catalysis in chemical reactions, chemical pressures or ionizations are some among several possible factors likely to influence characteristic properties of biophysical systems, including the equilibrium positions of protein molecules along macromolecular chains, their binding energies to the substrate and from a  general standpoint the shape of the double-well substrate. Biophysical systems with chain structures such as proteins cells, long DNA macromolecules, microtubules, polymers and biopolymers are relatively more flexible than solids, so taking into account this relatively high flexibility in their theoretical descriptions will undoubtedly provide a better understanding of their rich and peculiar dynamics.\par
\begin{figure}\centering
\begin{minipage}[c]{0.34\textwidth}
\includegraphics[width=2.in, height= 1.75in]{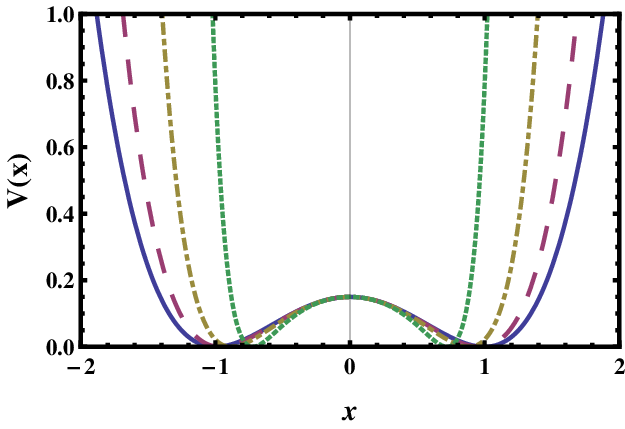}
\end{minipage}%
\begin{minipage}[c]{0.34\textwidth}
\includegraphics[width=2.in, height= 1.75in]{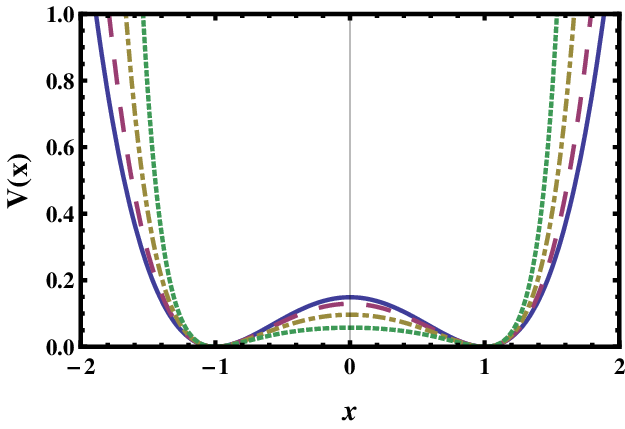}
\end{minipage}%
\begin{minipage}[c]{0.34\textwidth}
\includegraphics[width=2.in, height= 1.75in]{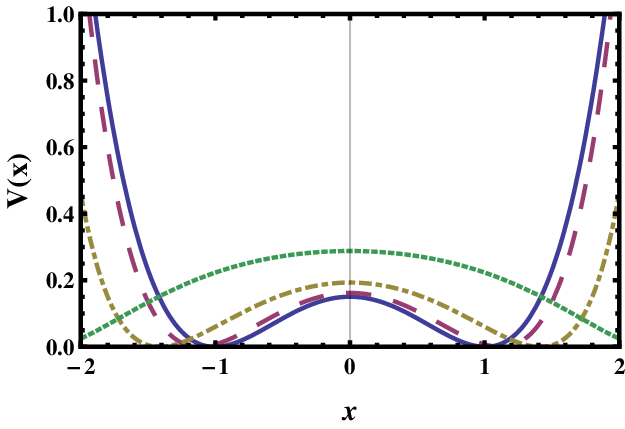}
\end{minipage}%
\caption{The three members of the family of deformable dougle-well potentials given by eq. (\ref{eq2}), plotted versus $x$ for $\mu=0.1$ (solid line)
   $\mu=0.5$ (dash line), $\mu=1.0$ (dash-dotted line), $\mu=2.0$ (dot line), $a_0=1.2$. Left graph is the DDWP I, middle graph is the DDWP II, right graph is the DDWP III.}
\label{fig:one}
\end{figure}
\begin{figure}\centering
\begin{minipage}[c]{0.52\textwidth}
\includegraphics[width=3.in, height= 2.25in]{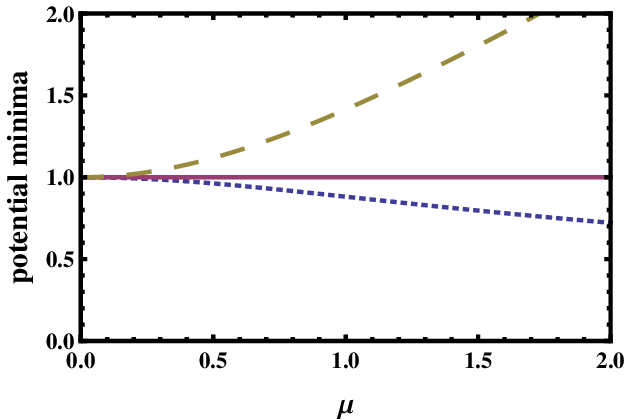}
\end{minipage}%
\begin{minipage}[c]{0.52\textwidth}
\includegraphics[width=3.in, height= 2.25in]{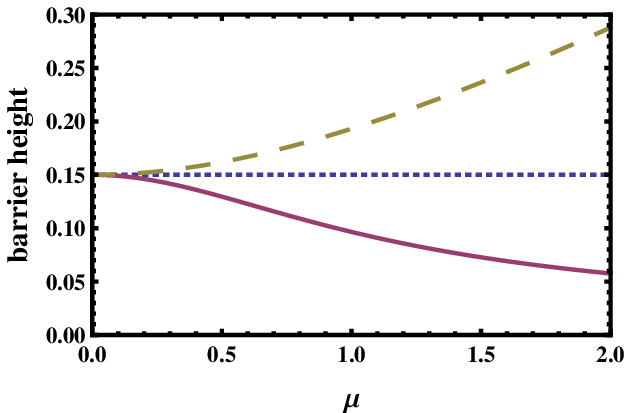}
\end{minipage}%
   \caption{Positive minima positions (left graph) and barrier height (right graph) of the three members of the family of DDWPs in eq. (\ref{eq2}), plotted versus $\mu$ for $a_0=1.2$. Dotted curve: DDWP I, solid line: DDWP II, dashed curve: DDWP III.}
   \label{fig:two}
\end{figure}

In the present study we consider the Kramers rate problem for bistable systems assuming a family of bistable potentials whose double-well shape can be tuned distinctly. This family of deformable double-well potentials (DDWPs) assumes the general form \cite{dik1,24}:
\begin{equation}
V_{\mu}(x)= \frac{a_0 q_{\mu}}{8}\left[\frac{1}{{\mu^2}}\sinh^2\left(\alpha_{\mu}x\right) - 1\right]^2, \hskip 0.35truecm \mu\neq 0, \label{eq2}
\end{equation}
where $\mu$ and $a_0$ are finite, real and positive. $q_{\mu}$ and $\alpha_{\mu}$ are two real parameters that can take three distinct forms depending on the specific feature of deformability of the double-well substrate. Below the three different deformable double-well potentials are denoted DDWP I, DDWP II and DDWP  III, in accordance with the following specifications:
\begin{enumerate}
 \item DDWP I:
 \begin{equation}
  q_{\mu}=1, \hskip 0.35truecm \alpha_{\mu}=\mu.
  \end{equation}
 In this case $V_{\mu}(x)$ is a double-well potential with fixed barrier height $V_b=V_{\mu}(0)$  but tunable minima positions \cite{dka}:
  \begin{equation}
  E_b= \frac{a_0}{8}, \hskip 0.5truecm  x_{1,2}=\pm \frac{arcsinh(\mu)}{\mu}.
  \end{equation}
 \item DDWP II:
 \begin{equation}
 q_{\mu}=\frac{\mu^2}{\left(1+\mu^2\right)\alpha^2_{\mu}}, \hskip 0.35truecm \alpha_{\mu}=arcsinh(\mu),
 \end{equation}
 $V_{\mu}(x)$ in this case is a double-well substrate whose barrier height $E_b$ can be tuned by varying $\mu$, leaving unaffected the two potential minima $x_{1,2}$ \cite{dkb}:
\begin{equation}
  E_b= \frac{a_0}{8}\frac{\mu^2}{(1+\mu^2)arcsinh^2(\mu)}, \hskip 0.5truecm  x_{1,2}=\pm 1.
  \end{equation}
 \item DDWP III:
 \begin{equation}
 q_{\mu}=\frac{\mu^2}{arcsinh^2(\mu)}, \hskip 0.35truecm \alpha_{\mu}=\frac{arcsinh(\mu)}{\sqrt{1+\mu^2}},
 \end{equation}
 $V_{\mu}(x)$ in this case is a double-well potential whose barrier height $E_b$ and minima positions $x_{1,2}$ are simultaneously shifted when $\mu$ is varied \cite{dkc}:
 \begin{equation}
 E_b=\frac{a_0}{8} \frac{\mu^2}{arcsinh^2(\mu)}, \hskip 0.5truecm x_{1,2}=\pm\sqrt{1+\mu^2}.
 \end{equation}
\end{enumerate}
When $\mu\rightarrow 0$ DDWP I, DDWP II and DDWP III reduce to the $\phi^4$ potential (\ref{eq1}). The three forms of DDWP are plotted in fig. \ref{fig:one} for four different values of the deformability parameter $\mu$ indicated in the caption. We also plotted the minima positions (fig. \ref{fig:two}, left graph) and barrier height (fig. \ref{fig:two}, right graph) of the three different forms of double-well potential $V_{\mu}(x)$, versus $\mu$ for $a_0=1.2$.
\par Our main objective in this study is to examine the impact of shape deformability on the escape rate of a Brownian particle moving in the field of force created by the double-well substrate (\ref{eq2}). In biophysical and chemical systems, the variations of characteristic parameters of the double-well substrate can result from several physical factors including chemical factors such as solvatation, catalysis and isotopic substitutions, or mechanical factors such as bond stretching or compression due to pressures, thermomechanical stresses from chemical reactions, etc. We derive the Kramers escape rate for the three double-well potentials taking into account non-Gaussian corrections.
\section{Kramers escape rate: general formulation and results for the DDWPs}
\begin{figure}\centering
\begin{minipage}[c]{0.34\textwidth}
\includegraphics[width=2.1in, height= 1.85in]{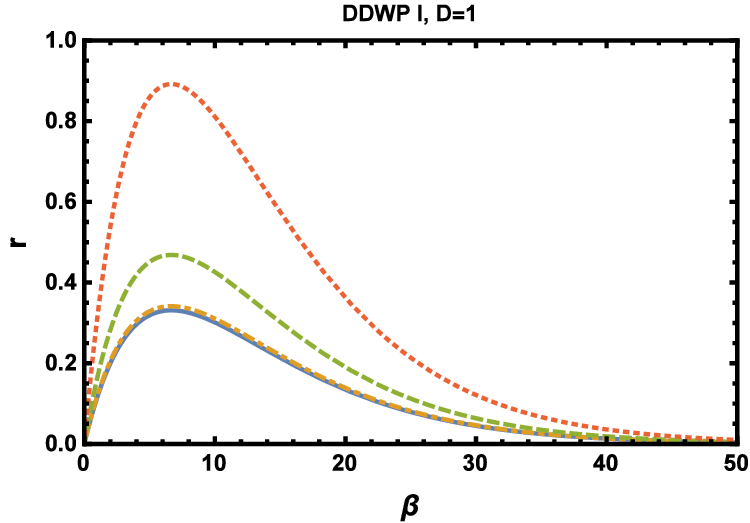}
\end{minipage}%
\begin{minipage}[c]{0.34\textwidth}
\includegraphics[width=2.1in, height= 1.85in]{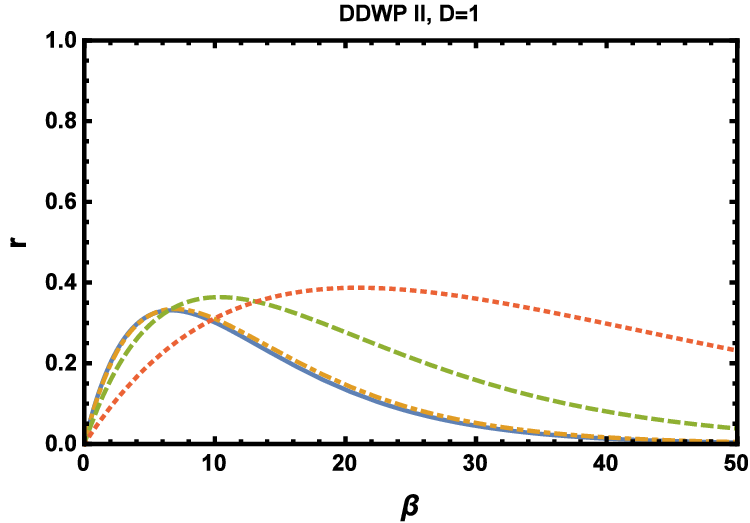}
\end{minipage}%
\begin{minipage}[c]{0.34\textwidth}
\includegraphics[width=2.1in, height= 1.85in]{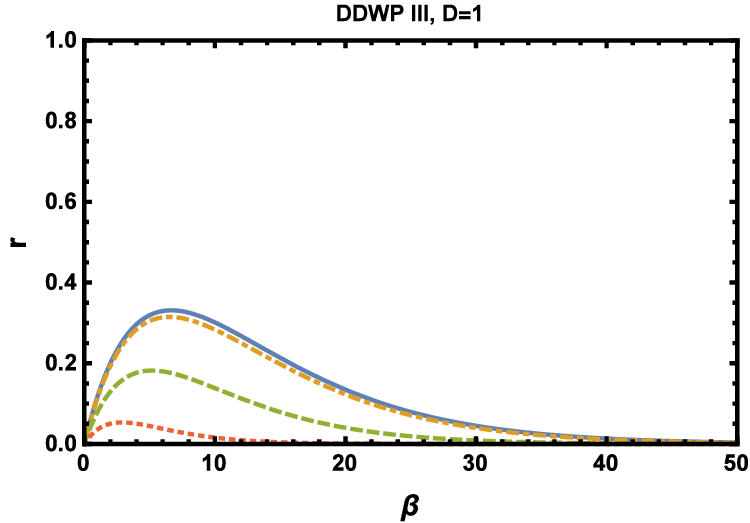}
\end{minipage}%
\caption{Non-corrected (i.e. Gaussian) escape rate versus $\beta$ for four different values of $\mu$ namely $\mu=10^{-6}$ (solid line), $\mu=0.25$ (dot-dashed line), $\mu=1$ (dashed line) and $\mu=2.5$ (dotted line). Left graph is for DDWP I, middle graph for DDWP II and right graph for DDWP III.}
\label{fig:three}
\end{figure}

Consider a collection of non-interacting identical Brownian particles of mass $m$, each placed in the field of force created by a double-well potential $V_{\mu}(x)$ where $x$ is the particle position at a given time $t$. In the overdamped limit, the ratio $m/\gamma$ where $\gamma$ is the coefficient of linear friction, is so small that the particle can be assumed unaccelerated. The dynamics of individual particles can therefore be approximated by the following Langevin equation:
\begin{equation}
\gamma \dot{x}=-\dfrac{\partial V_{\mu}(x)}{\partial x} + \eta(t),
 \label{eq3}
\end{equation}
where a dot symbol on top of a variable is a short-hand notation of its time derivative. The parameter $\eta$ is a random force with a Gaussian probability distribution and correlation function $<\eta(t)\eta(t')>=2\gamma D\delta(t'-t)$, with $\gamma D=1/\beta=K_BT$ where $D$ is the diffusion coefficient, $T$ is temperature and $K_B$ is the Boltzmann constant. The Fokker–Planck equation for the probability difstribution $P(x,t)$, corresponding to the Langevin equation (\ref{eq3}), is the Smoluchovski equation:
\begin{equation}
\frac{\partial}{\partial t}P(x,t) =\frac{\partial}{\partial x}\biggl(\frac{1}{\gamma}\frac{\partial}{\partial x}V_{\mu}(x) + D\frac{\partial}{\partial x}\biggr)P(x,t).
 \label{e3}                                   
\end{equation}
The Kramers problem for a bistable system is the one of first passage of a Brownian particle over a well of finite height $E_b$, moving from either of the two minima $x_1$ or $x_2$ of the double-well potential $V_{\mu}(x)$. To address the problem, we must determine the average time $\tau_{av}$ required for the particle to escape from
the potential well at $x_{1,2}$ across the wall at $x_0=0$. Two possible regimes of motion for the particle are envisageable: first the particle can remain trapped in an equilibrium state near the minimum where the energy is lowest, oscillating around this state at a characteristic frequency $\omega_m$ defined by:
\begin{equation}m\omega^2_m=\frac{\partial^2V_{\mu}(x)}{\partial x^2}\vert_{x=x_1}. \label{fr1}
\end{equation}
The second regime is the one where the particle is in a metastable state in between the minimum $x_1$ or $x_2$ and the maximum
$x_0=0$, but sufficiently far from $x_1$ or $x_2$. When the thermal energy $k_B T$ is far less than the barrier height $E_b$, the particle
will spend time visiting its equilibrium position at the potential bottom through forward-backward movements and only rarely will Brownian motions take it to the top of the barrier. However if the particle succeeds
to reach a position $x$ sufficiently close to the barrier top, it will be chanced to fall
rapidly to the other side and oscillate around the corresponding minimum for a while and then eventually cross back and resume oscillations around the
original minimum. Kramers postulated that the average time of first success of the particle in crossing the barrier $E_b$ at $x_0=0$ from the minimum $x_1$, is determined by the following equation:
\begin{equation}
\frac{\partial V_{\mu}}{\partial x}\frac{\partial \tau_{av}}{\partial x} - \beta^{-1}\frac{\partial^2 \tau_{av}}{\partial x^2}=\gamma. \label{kram1}
\end{equation}
To solve eq. (\ref{kram1}) we multiply by the exponential factor $\exp\lbrack-\beta V_{\mu}(x)\rbrack$ and integrate twice with respect to $x$, imposing that $\tau_{av}=0$ at $x=x_1$. This produces:
\begin{equation}
\tau_{av}=\frac{1}{D}\int_{x_1}^{x}{\exp\lbrack -\beta V_{\mu}(x')\rbrack dx' \int_{-\infty}^{x'}{\exp\lbrack \beta V_{\mu}(x")\rbrack dx"}} \label{kram2}.
\end{equation}
In Kramer's picture one requires that the exponential composing the first integrand is significant only near the minimum $x_1$, fading as we move away from the minimum. Hence we can expand $V_{\mu}(x')$ around $x_1$ and up to the quartic terms we get:
\begin{equation}
V_{\mu}(x')\approx \frac{m\omega_m^2}{2}(x'-x_1)^2 + \frac{\lambda_3}{3}(x'-x_1)^3 + \frac{\lambda_4}{4}(x'-x_1)^4 + 0(x'-x_1)^5. \label{ea}
\end{equation}
Similarly the exponential composing the second integrand in eq. (\ref{kram2}) must be significant only around the barrier top at $x_0=0$. So we can also expand $V_{\mu}(x")$ around $x_0$, and up to the quartic term we have:
\begin{equation}
V_{\mu}(x")\approx E_b - \frac{m\omega_0^2}{2} x"^2  + \frac{\zeta}{4} x"^4 + 0(x^5). \label{eb}
\end{equation}
Evidently the two exponential functions obtained from the approximations of $V_{\mu}(x)$ around $x_1$ and $x_0$, decay very fast with their arguments $x'-x_1=y$ and $x"-x_0=z$, respectively. Consequently it changes nothing quantitatively if we extend the finite integration boundaries to infinity. Doing this eq. (\ref{kram2}) becomes:
\begin{eqnarray}
\tau_{av}&=&\frac{I}{D}\exp(\beta E_b), \nonumber \\
I&=&\int_{-\infty}^{\infty}{\exp\lbrack -\beta m\omega_m^2y^2/2 -\beta\lambda_3y^3/3 -\beta \lambda_4y^4/4\rbrack dy \int_{-\infty}^{\infty}{\exp\lbrack -\beta m\omega_0^2z^2/2+\beta\zeta z^4/4\rbrack dz}}. \nonumber \\ \label{kram3}
\end{eqnarray}
By definition, the Kramers escape rate $r_{\mu}$ is the inverse of the average escape time $\tau_{av}$. Taking into account non-Gaussian corrections the Kramers escape rate is written:
\begin{eqnarray}
r^{nG}_{\mu}&=&1/\tau_{av} \nonumber \\
&=&\frac{D}{I}\exp(-\beta E_b). \label{kram4}
\end{eqnarray}
Let us evaluate the two non-Gaussian integrals in eq. (\ref{kram4}). In tables \ref{tab:one} and \ref{tab:two} we list the analytical expressions of coefficients in eqs. (\ref{ea}) and (\ref{eb}). Note that keeping only quadratic terms in the expansions of $V_{\mu}(x)$ around $x_1$ and $x_0$, the two integrals in (\ref{kram4}) are purely Gaussians i.e.:
\begin{equation}
\int_{-\infty}^{\infty}{\exp(-\beta m\omega_m^2y^2/2) dy \int_{-\infty}^{\infty}{\exp(-\beta m\omega_0^2z^2/2)}}=\frac{2\pi}{\beta m\omega_0\omega_m}, \label{kram5}
\end{equation}
leading to the following expression for the escape rate in the Gaussian regime:
\begin{equation}
r^{G}_{\mu}=r^0_{\mu}\exp(-\beta E_b), \hskip 0.5truecm r^0_{\mu}=\frac{\beta D m\omega_0\omega_m}{2\pi}. \label{kram6}
\end{equation}
\begin{table}
\caption{\label{tab:one} Analytical expressions of coefficients in eq. (\ref{ea}). }
\begin{ruledtabular}
\begin{tabular}{cccc}
Parameter&DDWP I & DDWP II & DDWP III\\
\hline
$\omega^2_m$&$a_0(1+\mu^2)/m$&$a_0/m$&$a_0/m$\\
$\lambda_3$&$3a_0(1+2\mu^2)\sqrt{1+\mu^2}$&$a_0\frac{3(1+2\mu^2)}{\mu\sqrt{1+\mu^2}}arcsinh(\mu)$ & $a_0\frac{3(1+2\mu^2)}{\mu(1+\mu^2)}arsinh(\mu)$ \\
$\lambda_4$
  &$a_0\lbrack 28\mu^2(1+\mu^2) +3\rbrack$
  &$a_0\frac{\lbrack 28\mu^2(1+\mu^2) +3\rbrack}{\mu^2(1+\mu^2)}arcsinh^2(\mu)$ & $a_0\frac{\lbrack 28\mu^2(1+\mu^2) +3\rbrack}{\mu^2(1+\mu^2)^2}arcsinh^2(\mu)$ \\
\end{tabular}
\end{ruledtabular}
\end{table}
\begin{table}
\caption{\label{tab:two} Analytical expressions of coefficients in eq. (\ref{eb}).}
\begin{ruledtabular}
\begin{tabular}{cccc}
Parameter&DDWP I & DDWP II & DDWP III\\
\hline
$Eb$&$\frac{a_0}{8}$&$\frac{a_0}{8}\frac{\mu^2}{(1+\mu^2)arcsinh^2(\mu)}$&$\frac{a_0}{8} \frac{\mu^2}{arcsinh^2(\mu)}$\\
$\omega^2_0$&$a_0/2m$&$a_0/2m(1+\mu^2)$&$a_0/2m(1+\mu^2)$ \\
$\zeta$
  &$a_0(3 - 2\mu^2)$
  &$a_0(3 - 2\mu^2)\frac{arcsinh^2(\mu)}{\mu^2(1+\mu^2)}$ & $a_0(3 - 2\mu^2)\frac{arcsinh^2(\mu)}{\mu^2(1+\mu^2)^2}$ \\
\end{tabular}
\end{ruledtabular}
\end{table}
In fig. \ref{fig:three} the non-corrected escape rate given by (\ref{kram6}) is plotted versus the inverse reduced temperature $\beta$, for $a_0$ and $D=1$. The left, middle and right graphs correspond respectively to DDWP I, DDWP II and DDWP III. The four curves in each graph are for four different values of $\mu$, selected arbitrarily to highlight the impact of the three distinct shape deformabilities of the double-well potentials, on their respective escape rate. As it is noticeable, the escape rate is more drastically affected by the deformability for the first (i.e. DDWP I) model. Indeed we observe an anhancement of the saturation pick of the escape rate for this model as the deformability parameter increases, suggesting that any physical process likely to shift the two degenerate minima positions away from the barrier position without affecting the barrier height, will favor the particle escale from its metastable state. Unlike the left graph in which the pick positions along the horizontal axis seem not to change when $\mu$ is varied, the middle and right graphs clearly show shifts in the temperature where the escape rate saturates before reentrance with decreasing temperature. The peak size decreases while being shifted toward zero values of $\beta$ as $\mu$ increases in the DDWP III model, while for DDWP II the escape rate is smaller for large $\mu$ in the large-temperature range, switching to larger values for large $\mu$ when $\beta$ is increased beyond a threshold value depending on $\mu$.
\par
\begin{figure}\centering
\begin{minipage}[c]{0.34\textwidth}
\includegraphics[width=2.1in, height= 1.85in]{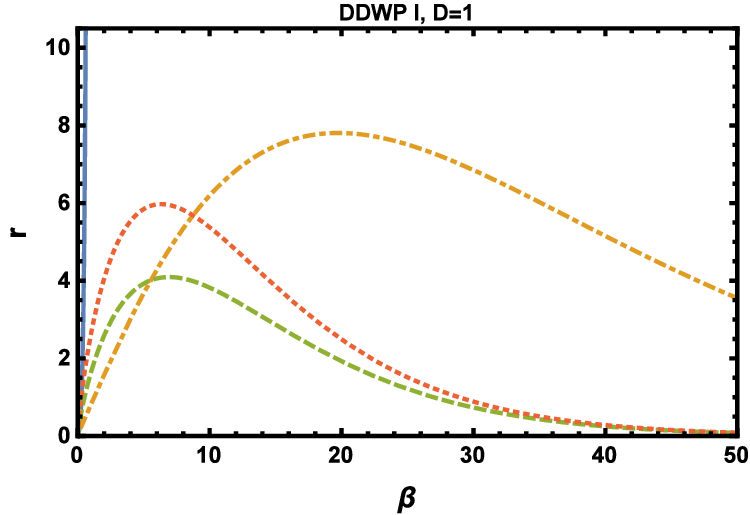}
\end{minipage}%
\begin{minipage}[c]{0.34\textwidth}
\includegraphics[width=2.1in, height= 1.85in]{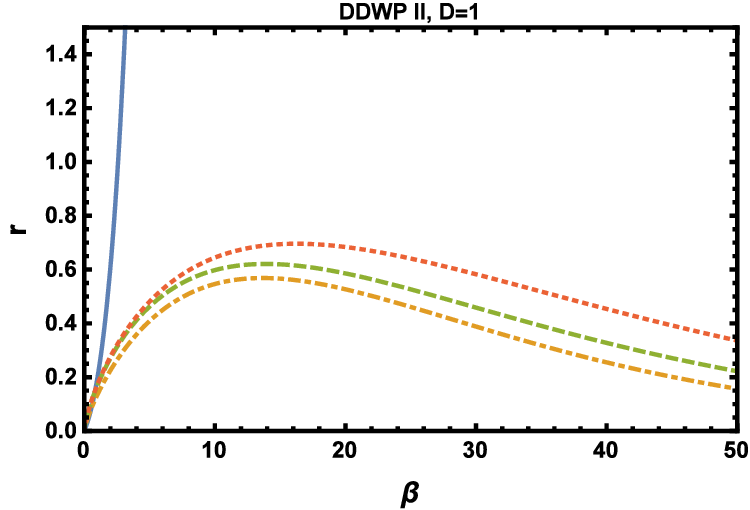}
\end{minipage}%
\begin{minipage}[c]{0.34\textwidth}
\includegraphics[width=2.1in, height= 1.85in]{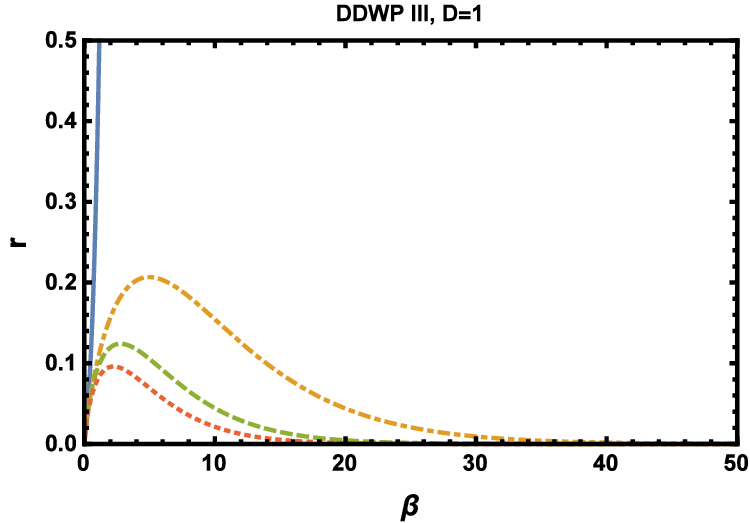}
\end{minipage}%
\caption{The corrected escape rate versus $\beta$ for four different values of $\mu$ namely $\mu=\sqrt{1.505}$ (solid line), $\mu=\sqrt{1.75}$ (dot-dashed line), $\mu=2$ (dashed line) and $\mu=2.5$ (dotted line). Left graph is for DDWP I, middle graph for DDWP II and right graph for DDWP III.}
\label{fig:four}
\end{figure}

 An observation of profiles of the three DDWPs shown in fig. \ref{fig:one} for some selected values of $\mu$, provides evidence that the shape of the potential around the minima $x_{1,2}$ remains dominantly sharp even when $\mu$ is given large values. The quadratic approximation is therefore reasonably acceptable for this situation, as opposed to the top of the barrier that seems to flatten as $\mu$ is increased, reminding us an inverted anharmonic quartic potential well. To this point in table \ref{tab:two}, the expression of the quartic coefficient $\zeta$ for the three forms of DDWP is manifestly sign changing with respect to the critical value $\mu_c=\sqrt{3/2}$ of the deformability parameter $\mu$. When $\mu<\mu_c$ $\zeta$ is positive and non-Gaussian corrections may be neglected, given that we are reasonably in the range of small values of $\mu$ for which the barrier top is sharp. When $\mu>\mu_c$, $\zeta$ becomes negative and the integral $I$ in eq. (\ref{kram3}) can be written:
\begin{equation}
I=\frac{1}{\omega_m}\sqrt{\frac{2\pi}{\beta m}}\,I_{nG}, \hskip 0.4truecm I_{nG}= \int_{-\infty}^{\infty}{\exp\lbrack -\frac{\beta m\omega_0^2}{2}z^2-\frac{\beta\vert\zeta\vert}{4} z^4\rbrack dz}.  \label{kram7}
\end{equation}
The non-Gaussian integral $I_{nG}$ is exact in terms of the real-argument parabolic cylinder function $\mathcal{D}_{-1/2}(y)$ \cite{wit};
\begin{equation}
I_{nG}= \biggl(\frac{2}{\beta\vert \zeta\vert}\biggr)^{1/4}\Gamma\lbrack\frac{1}{2}\rbrack\exp\biggl(\frac{\beta m^2\omega_0^4}{8\vert \zeta\vert}\biggr)\,\mathcal{D}_{-1/2}\biggl(m\omega_0^2\sqrt{\frac{2\beta}{\vert\zeta\vert}}\biggr).  \label{kram8}
\end{equation}
Substituting eq. (\ref{kram8}) in (\ref{kram4}) we obtain the the Kramers escape rate with non-Gaussian correction:
\begin{eqnarray}
r^{nG}_{\mu}&=&\tilde{r}_{\mu}\exp(-\beta E_b), \label{kram9}\\
\tilde{r}_{\mu}&=&\frac{D\omega_m}{\pi}\,\biggl(\frac{\beta\vert \zeta\vert}{2}\biggr)^{1/4}\,\sqrt{\frac{\beta m}{2}}\,\exp\biggl(-\frac{\beta m^2\omega_0^4}{8\vert \zeta\vert}\biggr)\mathcal{D}^{-1}_{-1/2}\biggl(m\omega_0^2\sqrt{\frac{2\beta}{\vert\zeta\vert}}\biggr). \label{kram9a}
\end{eqnarray}
The corrected escape rate is plotted in fig. \ref{fig:four} versus $\beta$, for four different values of $\mu$ chosen above the critical threshold $\mu_c$. Non-Guassian corrections are clearly seen to drastically enhance the escape rate in the three cases, with a relatively more pronounced impact on the DDWP I escape rate compared to the two others. Recall that eq. (\ref{kram9a}) was obtained on condition that $\mu>\mu_c$. When $\mu<\mu_c$ the argument of the parabolic cylinder function becomes complex and the escape rate loses its physical meaning.
As we discuss the implications of chosing $\mu$ with respect to a critical value $\mu_c$, it is interesting to mention our recent study \cite{dik1} of the issue of phase transition in quantum tunneling for systems with functional actions with the deformable double-well potential (\ref{eq2}). In that study we established that for $\mu<\sqrt{3/2}$ statistical mechanics predicts a second-order phase-transition in quantum tunneling (i.e. a second-order classical-to-quantum crosseover), consistent with Chudnovsky's criteria \cite{qpt} in the context of $\phi^4$ action. For values of $\mu$ above $\sqrt{3/2}$ a first-order classical-to-quantum crossover was observed, which turned out to be a distinctive signature of the model as a result of shape deformability of the double-well substrate.

\section{Summary and concluding remarks}
In this study we considered the Kramers rate problem for one-dimensional systems whose substrates are characterized by a bistable symmetry described by a deformable double-well potential. Systems of these kinds abund in nature, in general their structural orders involve two-state processes mediated by their environments. They are found in biophysics, biochemistry, chemistry, solid systems and most generally centrosymmetric systems. To proceed we considered a model describing a system of non-interacting
identical particles moving in the field of force created by a bistable substrate potential with a deformable double-well shape, in addition particles are subjected to frictions and a random force with Gaussian distribution. In the overdamped limit, where the dynamics of particles can be described by the Langevin equation, we formulated the equation of first-passage time for the Brownian particle moving from the bottom of the double-well potential. Three distinct forms of deformable double-well potentials were considered, characterized each by a specific deformability feature: in DDWP I the positions of the two degenerate minima are continuously shifted by tuning a single real parameter without affecting the barrier height, in DDWP II the barrier height can be shifted leaving unaffected the minima positions, and in DDWP III both the barrier height and minima positions can be tuned by varying the deformability parameter. The logic sustaining the choice of these three distinct forms of shape deformability resides in the possibility that each of them corresponds to a specific physical context enviseable in several biophysical, chemical and biochemichal contexts. To account for non-Gaussian corrections, the escape rate was obtained analytically by carrying out an expansion of the deformable-shape double-well potential beyond the Gaussian term as done in the standard Kramers rate theory \cite{hang1}. It turned out that non-Guaussian corrections become relevant for values of the deformability parameter above some critical threshold $\mu_c$, a critical value that already emerged in a previous work \cite{dik1} to set a border between two regimes of phase-transition in quantum tunneling: a second-order classical-to-quantum transition below $\mu_c$, and a first-order classical-to-quantum transition above $\mu_c$.\par The present study together with the recent one \cite{24} in which we constructed the spectral problem for instanton-phonon scatterings associated with the family of deformable double-well potentials eq. (\ref{eq2}), aims at highlighting the importance of taking into account the shape deformability in theoretical descriptions of physical processes involving flexible substrates with bistable symmetries.

\begin{acknowledgments}
The author is fellow of the Alexander von Humboldt Foundation.
\end{acknowledgments}

\section*{Author declarations}
\subsection*{Conflict of interest}
The author has no conflicts to disclose.
\subsection*{Author Contributions}
A. M. Dikand\'e: Conceptualization (equal); Formal analysis (equal); Investigation (equal); Methodology (equal); Resources (equal); Software (equal); Validation (equal); Writing-original draft(equal).
\subsection*{Data Availability}
Data sharing is not applicable to this article, as no new data were generated in the study.

\end{document}